\documentclass{article}
\usepackage[utf8]{inputenc}
\usepackage{amsmath}
\usepackage{amsthm}
\usepackage{amsfonts}
\usepackage{graphicx}
\usepackage{graphics}
\usepackage{authblk}
\usepackage{array}

\usepackage{amssymb}

\title{Efficient Rollup Batch Posting Strategy on Base Layer\thanks{We thank Lee Bousfield, Chris Buckland, Aysajan Eziz, Tim Roughgarden, and Thomas Thiery for valuable feedback.}}
\author{Akaki Mamageishvili}
\author{Edward W.  Felten}
\date{February 2023}
\affil{Offchain Labs}

\begin{document}

\maketitle

\begin{abstract}
    We design efficient and robust algorithms for the batch posting of rollup chain calldata on the base layer chain, using tools from operations research. We relate the costs of posting and delaying, by converting them to the same units and adding them up. The algorithm that keeps the average and maximum queued number of batches tolerable enough improves the posting costs of the trivial algorithm, which posts batches immediately when they are created, by $8\%$. On the other hand, the algorithm that only cares moderately about the batch queue length can improve the trivial algorithm posting costs by $29\%$. Our findings can be used by layer two projects that post data to the base layer at some regular rate.
\end{abstract}

\section{Introduction}

Ethereum blockchain satisfies high security and decentralization requirements, but it is not scalable. Namely, transaction fees are too high for running computationally heavy smart contracts. Also, it can only record a low number of transactions on each block. To improve scalability, several different solutions were proposed. Among them, one of the most successful solutions are layer two (shortly L2) networks, called rollup protocols. Several rollup protocols that build on top of Ethereum were proposed~\cite{arbitrum},~\cite{cartesi}. 
While the designs of these protocols of how heavy computation is delegated to them are different, they share one thing in common: from time to time they post (compressed) batches of transactions on the Ethereum main chain, referred to as a base layer or a layer one (L1). These recorded transactions on the base layer can be later used as a reference for rollup protocols and smart contracts governing them if there is a dispute about the state of execution of transactions.
Rollup protocols create new economic design questions to solve. In this paper, we address one of them. 
Namely, we study the question of posting data batches of layer two rollup chains on a base layer, which constitutes most of the costs the rollup chains incur. 
User transactions sent to the rollup protocol are grouped together with fixed frequency and compressed, which creates a batch. 
We study the question of how to decrease these costs in this paper. 
We try to find an efficient strategy for when to post transaction batches as the L1 price of posting data fluctuates. 
The trade-off is clear: avoid posting batches when the price is high, but at the same time, avoid delaying the posting. Such an algorithm adds to the robustness of the system, and overall, to its security. 

The question is motivated by historical experience with the Ethereum {\it base fee}, which fluctuates in a partially predictable way, and occasionally, when some major product is deployed on it, has intervals of very high base fee~\footnote{A memorable instance involved a base fee of more than 100 times the norm for a period of several hours. At least one rollup protocol (Arbitrum) decided to take manual control of the batch posting policy during that instance.}. Such intervals are not too long and not too frequent, which gives a hope that after some delay posting costs will be lower and posting can be resumed.

We decompose the cost in two parts: a posting cost that can be directly observed when posting batches, and a delay cost that is not directly measurable. More concretely, we interpret the total cost as the sum of posting and delay costs. Delay cost has several components. The first is psychological: users do not like when the batches are not posted for long period, as it may suggest to them that components of the system are down or unavailable. The second part is related to delayed finality: L2 transactions are not fully final until they are posted as part of a batch and that batch posting has finality on L1. Until that time, users must either wait for finality or trust the L2 system's sequencer to be honest and non-faulty. Delayed finality imposes costs on some applications.
The third part is related to the specific technical nature of transaction fee computation on L2 rollups. Namely, a transaction fee is calculated when transactions are created, not when they are posted on L1. Therefore, more delay causes less precise estimates of the L1 cost to attribute to L2 transactions, increasing the risk of unfair or inefficient pricing of rollup transactions. 

We model the problem as a Markov decision process. In each round, we calculate total costs independently from the past rounds. Each round is characterized by the current queue size and price, which constitutes a state. The price in the next round is modeled as a random variable, which depends on the current price. Depending on the strategy in the current round, the random variable indicating the price in the next round moves the state to the next state. For practical implementation, we discretize the price space and use a discrete approximation of a continuous random variable. To solve the optimization problem of finding the optimal strategy in each round, we use tools from dynamic programming, in particular, q-learning. The structure of the solution allows us to design a practical batch posting algorithm, which turns out to be intuitive and simple. The algorithm is characterized by only two parameters. We test the algorithm against a few natural benchmark algorithms, on the previous year's Ethereum base fee data~\footnote{The data covers the full period from the introduction of EIP-1559 in August 2021 until November 2022.}. Back-testing allows us to find optimal parameter sets for all algorithms and compare their performances. A comprehensive theoretical  analysis of the EIP-1559 gas fee scheme is given in~\cite{EIP1559}.

\subsection*{Related literature}

The optimization problem at hand is very similar to the inventory policy (IP) problem, long studied in economics and operations research literature, see~\cite{arrow_opt_inv} and~\cite{opt_inv_pol}. In IP problems, the newly produced items need to be sold for some price, and the demand price distribution is given. Therefore, the IP optimization problem is to maximize revenue, by maximizing revenue from trade and minimizing maintenance costs, while our problem is to minimize both costs. These two are dual. The only difference between our optimization problem and IP problems is that the delay cost in IP problems is linear in the inventory size, as the cost is interpreted as the maintenance cost of stored items, whereas we model the cost as superlinear in the number of delayed items. The optimality of pure stationary strategies in a wide range of Markov decision processes is shown in~\cite{stationary_optimal}. The main tool for solving our optimization problem is Q-Learning, which was introduced in~\cite{qlearning}. Linear delay cost of transaction inclusion is discussed in~\cite{bitcoin_delay}. The convergence of the dynamic programming algorithm that covers our case as well is discussed in~\cite{conv_qlearn}. 

\section{Model}

There is a discrete time with an infinite horizon. In round $i\in \mathbb{N}$, one new batch is created. The number of batches currently queued is denoted by $Q_i$. $P_i$ is the price of posting a batch in the round $i$.

At each round, the batch poster chooses a number of batches to publish, denoted by  $N_i$. It is proved in the operations research literature that the optimal way of doing so is to apply a stationary strategy. Namely, by applying a strategy function $S$ which must satisfy $0 \le S(P_i, Q_i) \le Q_i$. That is, it is optimal to assume that $N_i = S(P_i, Q_i)$ is only a function of the posting price in the current round and the queue size. By default, the algorithm posts batches that were created earliest (i.e., in FIFO order).  Intuitively, $S$ should be weakly increasing in $Q_i$ (if more batches to post, we should not post less) and  weakly decreasing in $P_i$ (if more expensive to post, we should not post more). This intuition can be used to test any (heuristic) solution we obtain. 

In round $i$, the system incurs a cost

\begin{equation}\label{cost_definition}
    C_i = C_{i,\mathrm{posting}} + C_{i, \mathrm{delay}} = P_i N_i + c(Q_i-N_i)^2.
\end{equation}
In this paper, we assume that $C_{i,\mathrm{posting}} := P_iN_i$ and $C_{i,\mathrm{delay}} := c(Q_i-N_i)^2$, where the $c>0$ coefficient represents the relative weight of the second component. 

The first term represents the cost of posting batches. Here we assume that the posting price is not affected by how many batches we post in each round. This assumption could be violated in practice if $N_i$ is very large. The second term represents the cost of delaying the posting of batches that remain in the queue after this round, essentially assuming that the cost of delaying a batch until the next round is linear in how long that batch has already been waiting~\footnote{Strictly speaking, the exact functional form of such delay would be $\sum_{i=1}^{Q_i-1}i=\frac{Q_i(Q_i-1)}{2}$, however, we approximate by ignoring the linear term in $Q_i$ and we absorb the factor of two into the coefficient $c$.}. The non-linear cost of delay is natural and is well-studied in the economics literature. 

\begin{figure}
    \centering
    \includegraphics[width = 10cm]{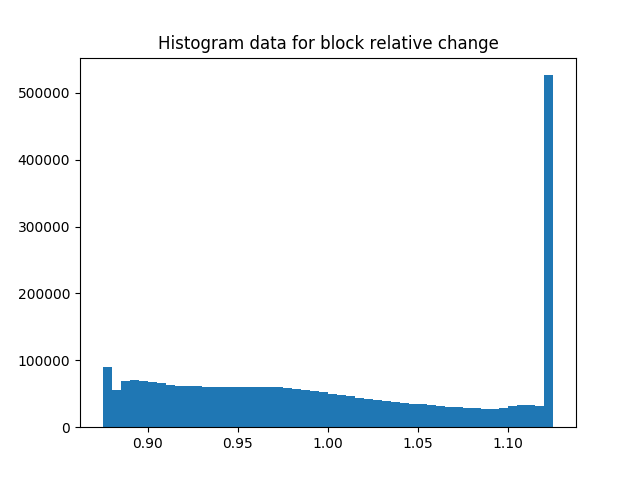}
    \caption{Relative change of the Ethereum base fee per block, over the past year. Each data point represents the ratio of the base fee in block $i+1$ to the base fee in block $i$. The large bar at 1.125 is the case where block $i+1$ is completely full so the base fee is increased by the maximum amount.}\label{fig:block_change}
\end{figure}

To move to the next round, we update: $$Q_{i+1}:= Q_i-N_i+1,$$ representing posting of $N_i$ batches in round $i$ and the arrival of one more batch in the queue in round $i+1$, and
$$P_{i+1} := R(P_i),$$
where $R$ is some random function that models the fluctuation of the batch posting price. Here we are making another implicit assumption that the L2 batch posting strategy does not affect the future base fee. This is a reasonable approach especially if the number of batches posted is not very large. Also, batches are created at a regular rate and their sizes are equal. The latter is relevant as the price of posting is measured in gas units, that is, the real cost is multiplied by how large the batches are in gas units.  

\begin{figure}\label{minute_change}
\includegraphics[width=10cm]{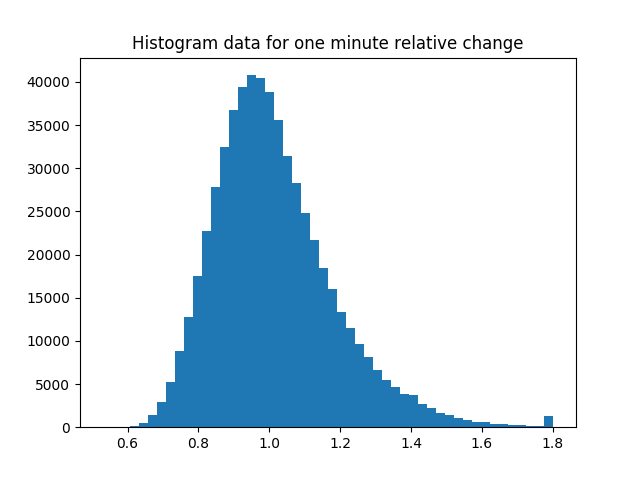}
    \caption{Relative change of the Ethereum base fee per minute, over the past year. Each data point represents the ratio of the base fee at $i+1$ minutes to the base fee at $i$ minutes. The distribution is similar to the result of composing five  steps drawn independently from the distribution in Figure~\ref{fig:block_change}.}\label{fig:minute-change}
\end{figure}

To begin, we consider the price of the next block to be uniformly distributed at the interval $[P_i\cdot{t_1},P_i\cdot t_2]$, where $t_1=\left(\frac{7}{8}\right)$ and $t_2=\left(\frac{9}{8}\right)$, as the base fee may change by $\frac{1}{8}$ either direction in every $12$ seconds. Note that $t_1\cdot t_2\approx 0.984$ is smaller than $1$, that is, the distribution is slightly skewed to the left. For a more realistic distribution, we need to look at the data. We assume that each batch is generated every minute (60 seconds), and the distribution of $R(P_i)$ is a result of uniform distribution convoluted $5$ times: $R(R(R(R(R(P_i)))))$. Theoretically, this approaches the normal distribution for a large enough number of convolutions. For data on the Ethereum one-minute base fee changes, see Figure~\ref{minute_change}. Note that there is a skew to the left direction, and there is an outlier at point $1.8$, which is caused by a large outlier in the block base fee change data, see Figure~\ref{fig:block_change}.
A large outlier corresponds to full blocks and a small outlier corresponds to empty blocks. Both outliers are observed in a theoretical setup with rational miners (block proposers), that maximize their own tips, see~\cite{farsighted}.

Although we do see a large peak at the max increase in the per-block data, we do not see a similarly large peak at the max increase in the per-minute data. This suggests that the per-minute data is likely more consistent with the hypothesis that per-block changes are independent of each other, and not that there are extended periods of the maximum price increase. 
Perhaps a large number of completely full blocks is due to different block producers having different lower bounds on the tip they require so that when a block producer with a low tip bound makes a block, that block includes a lot more transactions. That would be consistent with an assumption of independence between the base fee change in consecutive blocks. 

An especially interesting factor is a potential support size of $R(P_i)$. The right endpoint of the support turns out to be $1.8P_i$. The reason is that $1.8\approx (\frac{9}{8})^5$, and it seems to happen often that price increases exponentially~\footnote{Somewhat against the assumption of uniform distribution.}. In fact, if we look at block data, the maximum increase appears to happen $15\%$ of the time. The left endpoint is $0.6P_i$. The reason is that $0.6\approx (\frac{7}{8})^5$.    
    
\subsection*{Objective functions} 
Our goal is to minimize the expected value of the total cost. We assume there is an infinite horizon of rounds with discounted costs: we minimize $\sum_{i=0}^{\infty}C_i \delta^i$, where $\delta<1$ is a discount of future costs.
This variant is very similar to the inventory policy problem, with the only difference being that the delay cost in inventory policy problems is linear in the delayed number.

As a side note, we optimize the cost with a finite number of rounds $n$: In this variant, we minimize the total sum of costs $\sum_{i=1}^{n} C_i$. The solution is described in section~\ref{sc:finite_rounds}.

\section{Bellman Equation/Q-Learning}

We now turn to solve the main variant with an infinite number of rounds and future cost discounting.
To this end, we apply Q-Learning. Using standard notation\footnote{See Bellman equation: https://en.wikipedia.org/wiki/Q-learning.}, we calculate the following two matrices: $Q[s_t,a_t]$ and $O[s_t]$. $s_t$ is the current state and $a_t$ is an action. In our case, $s_t$ is a pair of $(Q_i,P_i)$, while $a_t$ is any natural number between 0 and $Q_i$. The action $a_t$ corresponds to how many batches to post at round $t$. $Q[s_t,a_t]$ denotes the total cost, discounting the future cost. $O[s_t]$ denotes the optimal action given the state $s_t$, that is, the action that minimizes the total cost from this point on. We initialize $O[s_t]$ with $Q_i,$ that is, the initial assumption that the optimal move is to publish all batches at each point. The value update iteration step is the following:

$$Q^{new}[s_t,a_t] := (1-\alpha) Q[s_t,a_t] + \alpha(c_t + \delta\cdot\mathbb{E}_R[\min_{a}Q[s_{t+1},a])].$$ 

$c_t$ is the cost incurred by taking action $a_t$, in this (stationary) round. That is, in our setting, this is $c_t:=a_t\cdot P_i+c(Q_i-a_t)^2$. $\alpha$ is a learning rate, as in the computation of the new matrix $Q$, we take the previously computed matrix $Q$ with weight $1-\alpha$ and new improved values with weight $\alpha$. After updating all states of the $Q$ matrix, we update all values of the $O$ matrix. Note that $O$  matrix values appear in the calculation of $\min_aQ[s_{t+1},a]$ which is replaced with $Q[s_{t+1},O[s_{t+1}]]$. 

We arbitrarily\footnote{Any other initialization works, for example, we can assign $0$ to all entries of $Q$ matrix. The only difference is in the convergence. While the rate stays the same, good initialization gives fewer iteration steps.} initialize $Q[s_t,a_t]$ as $a_t\cdot P_i + c(Q_i-a_t)^2$ and $O[s_t]$ as $Q_i$, consistent with an initial hypothesis, to be refined by Q-Learning, that the optimal move at each point in time is to publish all batches. 

\subsection{Implementation} 

We discretize the continuum price space by taking a bounded interval. That is, we assume that the price of batch posting will not go above some high bound. This is a reasonable assumption in our context.  
The space complexity of the implementation is $\Theta(PQ^2)$, where $P$ is the number of price points and $Q$ is the maximum number of batches we allow in the queue. To enforce this upper bound on queue length, we impose an infinite cost for exceeding the bound.

The run-time complexity of the system is $\Theta(IPQ^2f(P))$, where $I$ is the number of iterations before convergence, which depends on the convergence rate, which itself depends on the learning rate $\alpha$ and future discount $\delta$. $f(P)$ is an average of support sizes of $R(P_i)$ random variables.
In the case of a normal random variable, $f(P)\in \Theta(P)$. When generating a random variable for the close-to-boundary prices, we put weights only up until the upper bound price, as we do not have $Q$ matrix values. This irregularity of implementation creates misleading $O$ values for high prices, as the expectation by the implementation is that the price goes down in the next rounds, while theoretically, we would like to study a general case where there is no upper bound on the price. 

Higher $\delta$ means that we care about the future costs more. 
For the implementation, we take $\delta=0.999$, which is close enough to $1$, but not too close, as it slows down the computation considerably~\footnote{It also increases the magnitude of values in the matrix $Q$, as we weight future costs more.}. 

We take $P=400$ and $Q=300$. The ratio of the highest to lowest Ethereum base fee in our data is around $6000$. Therefore, to approximate the real price data with our discrete points, we multiply all prices by $\frac{6000}{400}=15.$ 
We take the learning rate $\alpha=0.1$. At this threshold, we observe that the values of the $Q$ matrix for large values of $P$ and $Q$ are stabilized. Generally speaking, lower $\alpha$ improves the precision of the $Q$ matrix calculation, but it increases computation time. Namely, the number of iterations is higher. In our case, we did not observe major qualitative or quantitative differences by choosing different values of $\alpha$.  
We assume that the algorithm has converged when the change in every entry in $Q$ is less than $\epsilon=0.01$, that is when: $$\max_{s_t,a_t}Q^{new}[s_t,a_t]-Q[s_t,a_t]<\epsilon.$$ The program takes $72$ hours to finish, for approximately 14000 iterations, on the input described above on Intel Core i7-8565U CPU @ 1.80GHz x 8. The current implementation is without parallelism. Note that $Q^{new}$ matrix calculation depends only on $Q$, therefore, full parallelization is feasible.  

\subsection{Observations}

Running the Q-learning algorithm and analyzing its output yields a few observations:

\begin{enumerate}
    \item There is a threshold price, below which all batches are posted. That is, there exists $T_P$, so that when $P_i<T_P$, $S_i(P_i,Q_i) = Q_i$.
    \item Above this threshold price, there exists a threshold on the number of batches that depends on the posting price, so that, below this threshold, no batches are posted. That is, there exists $T_Q(P_i)$, so that, when $Q_i<T_Q(P_i)$, $S_i(P_i,Q_i) = 0$. \item On the other hand, if $Q_i>T_Q(P_i)$, then $S_i(P_i,Q_i) = Q_i-T_Q(P_i)-1$. That is, the minimum number of batches is posted, to guarantee that in the next round if the price does not change, the threshold condition on the number of batches will still hold. 
\end{enumerate}

We conjecture that for any plausible functional form of the delay term of the cost, and for most smooth and convex/concave random distributions $R$ on price changes, these observations should hold. 
Verifying the hypothesis that this is the case for a given delay cost with a (smooth) version of random variable future batch posting price, by solving a dynamic programming problem for a given $R$, is left for the future.
The functional form of the function $T_Q(P)$ depends on the delay cost. For the quadratic delay cost, we conjecture that for $P > T_P$, the optimal threshold on the queue size, $T_Q(P)$, is of order $\Theta(\sqrt{P})$. 

\section{Back-testing results}

In this section, we compare the performance of the algorithm based on our Q-Learning analysis to three other algorithms' performances. We calculate how much each algorithm would spend and how much delay cost each would incur, on the last year's time-series data of Ethereum base fees, taken after every minute, that is, after every fifth block\footnote{Before the Ethereum Merge on 6 September 2022, the time between blocks was about 12 seconds on average. Since the Merge, the interval between blocks is fixed at 12 seconds.}. It is assumed that each batch has the same unit size, which is a good approximation in practice. If batches are created with less frequency, we can easily modify the test set and optimize parameters accordingly. 

For each algorithm, we measure a few properties: publishing cost; delay cost; the average and worst-case delays experienced by batches; and the maximum number of batches posted in any one round. The last measure is an important robustness measure, as posting too many batches at the same time may affect the future price or even be impossible to perform given the L1 block space constraints. All performances given in the following subsections are on the Pareto-efficient curve of the pair of publishing and delay costs. Back-testing algorithms were optimized for approximately $100$ different instances. All algorithms described below are linear in the data size, as the decision in each round depends only on the current round price and queue size.      

\subsection{Current Arbitrum algorithm} The first algorithm is currently deployed by Arbitrum. We denote it by $\mathcal{O}$. $\mathcal{O}$ is characterized by 3 parameters, over which we minimize its total cost:

\begin{itemize}
    \item $ap$, intuitively an {\it acceptable price} measured in GWEIs,
    \item $e$, an exponent,
    \item and $ut$, an {\it update time} in minutes.
\end{itemize}

Every batch has these three features. If the base fee is lower or equal to $ap$, the batch is posted. After $ut$ time passes and the batch is still not posted on L1, the new acceptable price becomes $e$ times bigger. That is, $ap^{new}:=e\cdot ap$.

The performance of the algorithm $\mathcal{O}$ for a few parameter sets is documented in the following table. 
\begin{center}
\begin{tabular}{ | m{7em} | m{2cm}| m{2cm}|   m{2cm}| m{2cm}|}
  \hline
  Parameters $(e, ap, ut)$ & Publishing cost & Delay cost & Maximum delay & Avg. delay \\ 
  \hline
  $(1.2, 144, 60)$ & $2.598e+07$ & $5.99e+09$ & $773$ & $39.7986$ \\ 
  \hline
  $(2, 96, 120)$ & $3.170e+07$ & $6.456e+07$ & $193$ & $2.23403$ \\ 
  \hline
  $(2.8, 72, 140)$ & $3.271e+07$ & $1.866e+07$ & $165$ & $0.949142$ \\
  \hline 
\end{tabular}
\end{center}

Note that publishing cost increases and maximum delay and average delay decrease by increasing $e$. Both behaviors are natural as the algorithm posts more aggressively for higher $e$. There is no clear upper bound on the delay in round $i$ as a function of posting price $P_i$, as it depends on when the exponential function catches up with the price, i.e., it depends on the history as well.  

The following table shows the performance of the Arbitrum algorithm where the acceptable price does not do a step-function doubling every update time but instead increases in a smooth exponential curve with the same doubling time. That is, the exponent in each round is equal to $e^{\frac{1}{ut}}$. 

\begin{center}
\begin{tabular}{ | m{7em} | m{2cm}| m{2cm}|   m{2cm}| m{2cm}|}
  \hline
  Parameters $(e, ap, ut)$ & Publishing cost & Delay cost & Maximum delay & Avg. delay \\ 
  \hline
  $(1.2, 144, 60)$ & $2.60294e+07$ & $4.73071e+09$ & $771$ & $33.5433$ \\ 
  \hline
  $(2, 96, 120)$ & $3.19334e+07$ & $1.72324e+07$ & $183$ & $1.08583$ \\ 
  \hline
  $(2.8, 72, 140)$ & $3.31778e+07$ & $4.86956e+06$ & $164$ & $0.429163$ \\
  \hline 
\end{tabular}
\end{center}

Compared to the step algorithm described above, publishing costs are slightly increasing and delay costs are slightly decreasing. But qualitatively, the results are very similar. 

\subsection{Q-learning algorithm} 
The second algorithm, $\mathcal{Q}$, is based on Q-Learning.  
We optimize over two parameters: $d$ and $T_p$, when minimizing the cost of the $\mathcal{Q}$ algorithm. In particular, we test $T_Q(P)=\frac{\sqrt{P-T_P}}{d}$ and $T_P := T_p$.  

The performance of the algorithm $\mathcal{Q}$ for a few parameter sets is documented in the following table. 

\begin{center}
\begin{tabular}{ | m{5em} | m{2cm}| m{2cm}|   m{1.6cm}| m{1.7cm}| m{1.3cm}|}
  \hline
  Parameters $(T_p,d)$ & Publishing cost & Delay cost & Max. delay & Avg. delay & Max. posted \\ 
  \hline
  $(60, 2)$ & $3.420e+07$ & $1.283e+06$ & $42$ & $1.699$ & $5$ \\ 
  \hline
  $(60, 1.6)$ & $3.383e+07$ & $2.339e+06$ & $52$ & $2.127$ & $7$  \\ 
  \hline
  $(80, 1.2)$ & $3.324e+07$ & $4.111e+06$ & $69$ & $2.00$ & $10$  \\
  \hline 
\end{tabular}
\end{center}

Note that both delay cost and maximum delay are increasing in decreasing $d$, while publishing cost is decreasing. 
The publishing cost is decreasing and the delay cost is increasing with increasing $T_P$. As a robust measure for $T_P$, we can take the value that is $80\%$ percentile of the base fee data distribution. This allows the algorithm to be fully automatic, by updating $T_P$ every month or two weeks, to adjust to the current trend of prices. 
We do not include a maximum posted number in the performance of the other algorithms as this number is equal to the maximum delay by definition. Unlike the previous algorithm, the algorithm in this section gives an upper bound on the delay in each round $i$ as a function of the price of posting $P_{i}$, which is equal to $\frac{\sqrt{P_{i}-T_P}}{d}$ and adds to the robustness of the system. This gives a global upper bound $\frac{\sqrt{P_{\max}-T_P}}{d}$ on the maximum delay, where $P_{\max}$ is the maximum posting price. Note that the maximum base fee was about $8200$ GWEIs and the maximum delays in the table approximately correspond to this upper bound. For example, $\frac{\sqrt{8200}}{1.2}\approx 74 > 69$. In general, the upper bound does not have to be achieved, as the price may increase very quickly and the delay queue size may not catch up.

\subsection{Price minimizing algorithm}
The third algorithm, $\mathcal{D}$, delays posting until the price drops below a certain threshold $T$ and then posts all batches.

The performance of the algorithm $\mathcal{D}$ for a few parameters is documented in the following table. 
\begin{center}
\begin{tabular}{ | m{5em} | m{2cm}| m{2cm}|   m{2cm}| m{2cm}|}
  \hline
  Parameters $T$ & Publishing cost & Delay cost & Maximum delay & Avg. delay \\ 
  \hline
  $60$ & $2.237e+07$ & $2.658e+12$ & $14975$ & $584.191$ \\ 
  \hline
  $80$ & $2.632e+07$ & $2.683e+11$ & $8448$ & $124.4$ \\ 
  \hline
  $100$ & $2.896e+07$ & $3.835e+09$ & $1394$ & $15.66$ \\
  \hline 
\end{tabular}
\end{center}

Notice that a huge value of the maximum delay would violate our assumption that the price of posting in each round is not affected by how many batches we post. We do not include a maximum posted number in the performance of the other algorithms as this number is equal to the maximum delay by definition.  



\subsection{Trivial algorithm}

A trivial algorithm, denoted by $\mathcal{T}$, posts every batch immediately.
$\mathcal{T}$ has its own merits. First, it keeps the delay costs to $0$, which imposes a minimal load on the system. Second, it is simple to interpret and easy to implement.

Testing on the same data as the above algorithms, we find that a trivial algorithm has publishing costs equal to $3.6e+07$. Note that it does not have any parameters and the delay cost is equal to $0$. We use this result as a benchmark to measure the performance of other algorithms with respect to the cost of publishing.  

\subsection{Tips}
In the paper so far, we ignored tips that are given to L1 block builders for inclusion in the block in the design and analysis of efficient batch posting strategy. These tips should in principle be counted towards the price of publishing. For example, Arbitrum has a fixed tip, 1 GWEI per gas, which is enough to get included in $95\%$ of the cases. If we count the minimum tips to be included in each block towards the total price of being posted and run the same algorithms as before, we get very similar results as before, namely, in $1\%$ proximity of both, publishing and delay costs. One potential explanation is that when tips to be included are high, base fees are also high, therefore, none of the algorithms post their batches. These observations can be seen as indicators to simplify the decision problem by including the tips directly towards the cost of publishing, instead of choosing them strategically.  

\section{Dynamic programming with fixed prices}
\label{sc:finite_rounds}
In this section, we discuss the case where there is a fixed number of rounds $n$, and posting prices in each round are fixed and given in advance. At the end of the last round, we publish all batches that are left unpublished. The use-case of this algorithm, for example, is if there are futures contracts on base fees.  

The optimum solution can be found by dynamic programming with run-time $\Theta(n^3)$ as described in the following. In $dp[i][j]$, we store the minimum cost incurred if in the first $i$ rounds we publish exactly $j$ batches. We iterate $i$ through all rounds in the outermost loop, contributing the first multiplicative factor $n$. In the second loop, we iterate $j$ between 0 and $i$, contributing another multiplicative factor $n$. In the third and innermost loop, we iterate $take$ over newly published batches, between 0 and $i+1-j$, therefore, contributing the last multiplicative factor $n$. We update $dp[i+1][take+j]$ with the maximum between the following two values: 
\begin{equation}
    dp[i+1][take+j] := \max(dp[i+1][take+j], dp[i][j]+cost),
\end{equation} where $cost$ is calculated as $c(i-j-take+1)^2.$ We also record $take$ that gives the minimum answer for each $i$ and $j$, which will allow us to recover the answer, and the number of batches published at each round to minimize global cost. The global cost is located at $dp[n][n]$ and we can reconstruct the answer of how many to publish at each round using a backtracking algorithm. 

We generated prices according to different distribution functions and observed that it is almost always optimal to publish zero or all batches.

\section{Conclusions}

We initiate the study of an efficient batch posting strategy by L2 rollup chains on the L1 chain as a calldata. As an outcome, we obtain efficient algorithms with robustness guarantees. Namely, in each round, the new algorithm does not post too many batches and the number of batches kept in the queue is bounded by a function of posting price in each round. Future avenues of research include the optimization problem where current and future prices depend on the number of batches posted in each round. This may be the case if rollup protocols become dominant in the scalability of the base fee. Finding out the optimal constant tip is also left for future research.

\bibliographystyle{plain}
\bibliography{sample}

\begin{thebibliography}{10}

\bibitem{arrow_opt_inv}
Kenneth~J. Arrow, Theodore Harris, and Jacob Marschak.
\newblock Optimal inventory policy.
\newblock {\em Econometrica}, 19(3):250--272, 1951.

\bibitem{stationary_optimal}
Hugo Gimbert.
\newblock Pure stationary optimal strategies in markov decision processes.
\newblock In Wolfgang Thomas and Pascal Weil, editors, {\em {STACS} 2007, 24th
  Annual Symposium on Theoretical Aspects of Computer Science, Aachen, Germany,
  February 22-24, 2007, Proceedings}, volume 4393 of {\em Lecture Notes in
  Computer Science}, pages 200--211. Springer, 2007.

\bibitem{farsighted}
Jens~Leth Hougaard and Mohsen Pourpouneh.
\newblock Farsighted miners under transaction fee mechanism {EIP1559}.
\newblock {\em Working Paper}, 2022.

\bibitem{bitcoin_delay}
Gur Huberman, Jacob~D. Leshno, and Ciamac Moallemi.
\newblock Monopoly without a monopolist: An economic analysis of the bitcoin
  payment system.
\newblock {\em The Review of Economic Studies}, 88(6):3011--3040, 2021.

\bibitem{conv_qlearn}
Tommi~S. Jaakkola, Michael~I. Jordan, and Satinder~P. Singh.
\newblock On the convergence of stochastic iterative dynamic programming
  algorithms.
\newblock {\em Neural Comput.}, 6(6):1185--1201, 1994.

\bibitem{opt_inv_pol}
Arthur F.~Veinott Jr.
\newblock The optimal inventory policy for batch ordering.
\newblock {\em Operations Research}, 13:424--432, 1965.

\bibitem{arbitrum}
Harry~A. Kalodner, Steven Goldfeder, Xiaoqi Chen, S.~Matthew Weinberg, and
  Edward~W. Felten.
\newblock Arbitrum: Scalable, private smart contracts.
\newblock In William Enck and Adrienne~Porter Felt, editors, {\em 27th {USENIX}
  Security Symposium, {USENIX} Security 2018, Baltimore, MD, USA, August 15-17,
  2018}, pages 1353--1370. {USENIX} Association, 2018.

\bibitem{EIP1559}
Stefanos Leonardos, Barnab{\'{e}} Monnot, Dani{\"{e}}l Reijsbergen, Efstratios
  Skoulakis, and Georgios Piliouras.
\newblock Dynamical analysis of the {EIP-1559} ethereum fee market.
\newblock In Foteini Baldimtsi and Tim Roughgarden, editors, {\em {AFT} '21:
  3rd {ACM} Conference on Advances in Financial Technologies, Arlington,
  Virginia, USA, September 26 - 28, 2021}, pages 114--126. {ACM}, 2021.

\bibitem{cartesi}
Diego Nehab and Augusto Teixeira.
\newblock The core of cartesi.
\newblock {\em White Paper}, 2018.

\bibitem{qlearning}
Christopher John Cornish~Hellaby Watkins.
\newblock {\em Learning from delayed rewards}.
\newblock 1989.

\end{thebibliography}

\end{document}